\documentstyle[12pt]{article}

\begin{document}
\begin{titlepage}
\begin{flushright}
\vbox{EHU-FT-94/10\\
hep-th/9412003
}
\end{flushright}
\vskip 5cm
\begin{center}
{\Large\bf Remarks on the Atick-Witten behavior and strings
near black hole horizons}
\vskip 1cm
R.\ Emparan\\
{\it
Depto. F{\'\i}sica de la Materia Condensada\\
Universidad del Pa{\'\i}s Vasco\\
Apdo.\ 644, 48080 Bilbao, SPAIN
}
\\
{\tt
wmbemgar@lg.ehu.es}
\end{center}
\date{November 1994}
\vskip 1.5cm
\begin{abstract}
We present arguments pointing to a behavior
of the string free energy
in the presence
of a black hole horizon similar to
the Atick-Witten dependence on temperature beyond the Hagedorn
transition. We give some evidence based on orbifold
techniques applied to Rindler space and further support is found
within a Hamiltonian approach. However, we argue that
the interpretation in terms of a
reduction of degrees of freedom is confronted by serious
problems. Finally, we point out the problems concerning heuristic
red-shift arguments and the local interpretation of
thermodynamical quantities.
\end{abstract}
\end{titlepage}
In attempts to better understand the nature of
string theory, the behavior of strings
at high energies \cite{gros} and high temperatures \cite{ati}
has been investigated in recent years. In these regimes the
fundamental degrees of freedom of the string are expected to
show up and a recurrent feature in these investigations has been
the fact that string theory at high energies seems to have a
surprisingly low number of fundamental degrees of freedom, far
fewer than expected in relativistic field theories.

There is still another scenario where the fundamental degrees
of freedom of the string are expected to play a prominent role,
i.e., the horizon of a black hole. Pioneering investigations by
't Hooft \cite{thoo}Êshowed that thermodynamical quantities like
the entropy or the free energy of quantum fields diverge near
the horizon. Since the presence of the horizon involves
arbitrarily high frequencies, it is argued that a
proper description must make reference to ultra-short distance
physics where string theory is expected to play a fundamental
role.

We will investigate whether the results of
Atick and Witten \cite{ati}, which showed that the high
temperature behaviour of the free energy of the string
presents a dependence on the temperature characteristic of two
dimensional field theories, can be extended in the presence of
a black hole horizon. Due to technical difficulties in
quantizing the string in black hole backgrounds, the
conclusions will not be completely air-tight, but different
approaches will coincide in giving evidence of an
Atick-Witten behavior. However, for reasons discussed below the
statistical interpretation is not straightforward.

We start by briefly reviewing the analysis in Ref.\
\cite{ati}. The starting point is the closed bosonic
string free energy \cite{pol,obr}
\begin{equation}\label{modf}
{}F(\beta)=-{V\over 4\pi^2\alpha'}\int_{\cal F}
{d^2\tau\over\tau_2^2}
\lambda(\tau){\sum_{m,n}}'\exp(-{\beta^2|m\tau+n|^2\over
4\pi\alpha'\tau_2})
\end{equation}
where $\lambda(\tau)$
is a modular invariant quantity given by
\begin{equation}\label{lambd}
\lambda(\tau)={|\eta(\tau)|^{-48}\over
(4\pi^2\alpha'\tau_2)^{12}}
\end{equation}
and the integration region ${\cal F}$ is the fundamental domain
of the torus moduli space,
$-1/2<\tau_1<1/2,\, |\tau|>
1,\, \tau_2>0$,
which excludes the ultraviolet region $\tau_2\sim 0$. The term
$m=n=0$ corresponding to the vacuum free energy is excluded
from the sum.

The behavior for $\beta\rightarrow 0^+$ was
analyzed in Ref.\ \cite{ati} by noting that, in that limit, the
sums can be replaced by integrals,
\begin{equation}\label{sumint}
\sum_{m,n}\rightarrow \beta^{-2}\int_{-\infty}^\infty dy_1 dy_2
\end{equation}
with $y_1=\beta m$ and $y_2=\beta n$.
One then easily finds
\begin{equation}\label{f2}
\beta F\simeq {V\over \beta}4\pi^2\alpha'\Lambda
\end{equation}
where $\Lambda$ is the one loop cosmological constant of the
closed bosonic string. The latter is ill defined due to the
presence of a tachyonic infrared divergence. A similar
divergence is still present for the Type II superstring,
showing that introduction of supersymmetry does not solve the
problem. The origin of these divergences can be easily traced:
the free energy Eq.\ (\ref{modf}) is ill defined beyond a
certain value of $\beta$ due to the so-called Hagedorn
singularity in the region $\tau_2\rightarrow\infty$
\cite{obr,ati}. By examining the leading exponential behavior
in this limit, we find an infrared
divergence of the modular integral when
$\beta\leq
4\pi\sqrt{\alpha'}\equiv \beta_c$ \cite{obr}. This is the
Hagedorn instability.

Notice that the temperature dependence in Eq.\ (\ref{f2})
followed from simple formal power counting. Since a thermal
dependence of this kind is what one would expect from a
field theory in two dimensions, Atick and Witten
have argued that, in spite of infrared singularities, this
result can be interpreted to indicate a vast reduction of the
fundamental degrees of freedom in string theory.

We now turn to study whether a similar behavior occurs in
the presence of a black hole horizon. Our analysis will be
carried out in Rindler space, which approximates the near
vecinity of non-extremal black hole horizons, or equivalently,
the geometry outside very large black holes (with vanishingly
small curvature).
The euclidean metric of Rindler space is
\begin{equation}\label{rindmet}
ds^2=\xi^2 d\theta^2+d\xi^2+\sum_{i=1}^{D-2} dx_i^2\,.
\end{equation}
Here $\theta$ is the euclidean time to be periodically
identified $\theta\sim\theta+\beta$, so that for $\beta\neq
2\pi$ a conical singularity is present. Sections $\xi={\rm
const.}$ correspond to observers undergoing constant acceleration
$a=1/\xi$ or, in the black hole picture, at fixed distance above
the horizon, which is located at $\xi=0$. The regularity
requirement that $\theta\sim\theta+2\pi$ corresponds to choosing
the Hartle-Hawking thermal vacuum, which, in this approximation,
is the Minkowski vacuum. It is the Hartle-Hawking state which
corresponds to thermodynamic equilibrium with the (eternal)
black hole. However, thermodynamical analysis requires
considering general values $\beta\neq 2\pi$, as is clear, for
instance, from the formula for the entropy
\begin{equation}\label{entf}
S=(\beta{\partial\over\partial\beta}-1)\beta F
\end{equation}

Now, quantization of string theory in the Rindler background
with $\beta\neq 2\pi$ is far from straightforward. If we
want to compute the free energy within an euclidean functional
integral approach, we find that naive direct introduction of a
conical singularity, e.g. by introducing a singular curvature
term in the nonlinear sigma model, spoils conformal invariance.
Consistent conical backgrounds for strings are obtained in the
form of orbifolds \cite{dix} and they have been applied to this
problem in Refs.\ \cite{dab,low}.

The main distinguishing feature of the free energy in $Z_N$
orbifold string theory, as developed in Refs.\
\cite{dab,low}, is a
double sum over twists in the nontrivial cycles of
the torus corresponding to the fact that the winding number
around the cone apex is conserved only modulo $N$, where
$N=2\pi/\beta$. Recall
that the double sum over solitons in Eq.\ (\ref{modf}) was
responsible for the high temperature dependence of the free
energy. Remarkably, due to the double sum over
twists (and after taking into account an additional
factor $1/N$ coming from projection over $Z_N$-invariant
states) one can expect a leading dependence of the one loop
string amplitude like $A_N=-\beta F\sim N\sim \beta^{-1}$. In
fact, in Ref.\ \cite{dab} it has been noticed that, in the large
$N$ limit, the one loop string amplitude
$A_N=-\beta F$ behaves like $\sim N\log N$, i.e., $\sim
\beta^{-1}\log \beta$, where the leading power factor comes from
the double sum over twists. We want to point out the (previously
unnoticed) similarity of this leading power dependence to the
Atick-Witten behavior. It is
likely that this dependence will hold when analytically
continuing to non-integer $N$, but the discrete character of
the sums makes it difficult to establish more precisely the
relation and we will have to look for further evidence from
other approaches.

We will follow now a more pedestrian Hamiltonian treatment. Since
the current algebra relations that determine the mass spectrum
are invariant under field redefinitions, we expect
the string spectrum in the Rindler background
to be the same as in flat space. There could be a problem here
if we wanted to regularize the horizon by cutting off a region
$\xi<\xi_0$, because this introduces highly nontrivial
complications (see Ref.\ \cite{bar2} for a discussion). However,
in the orbifold method one works in the whole conical space
without introducing any cutoff. Therefore we
will extend our discussion to distances arbitrarily close to the
horizon $\xi=0$, a way to proceed analogous to the limit
$\beta\rightarrow 0^+$ considered in Ref.\ \cite{ati}, and we
will compute stringy quantities by summing over the already known
string spectrum in flat space.

We are thus led to computing the free energy in a conical
background. This can be conveniently performed by
using heat kernel methods to write
\begin{equation}\label{hkfree}
{}F(\beta)=-{1\over 2\beta}\int_0^\infty
{ds\over s}\zeta(s)e^{-s m^2}
\end{equation}
where $\zeta(s)$ is the heat kernel of the Laplacian in the
Rindler geometry, Eq.\ (\ref{rindmet}). As argued above, we will
obtain the string free energy by summing Eq.\
(\ref{hkfree}) over the string spectrum of oscillators. This is
given by the mass formula
\begin{equation}
{\alpha'\over 2} m^2=-2+N+\tilde N
\end{equation}
where $N$ and $\tilde N$ are left and right moving mode
number operators. Additionally, the restriction $N=\tilde N$
must be imposed.
The oscillator sum can be readily performed \cite{pol} and yields
\begin{equation}\label{hkfreest}
{}F(\beta)=-{1\over 2\beta}\int_{-1/2}^{1/2}
d\tau_1\int_0^\infty {ds\over s}\zeta(s)
|\eta(\tau)|^{-48}
\end{equation}
with
$\tau=\tau_1+i\tau_2,\, \tau_2\equiv{s/\pi\alpha'}$.
Here, the integral over $\tau_1$ is introduced as a means
to implement the constraint $N=\tilde N$, and the dimension has
been set to $D=26$.

In the Rindler geometry of Eq.\ (\ref{rindmet}), the heat kernel
$\zeta(s)$ factorizes into contributions from the $(D-2)$
dimensional transverse flat space and the two dimensional cone.
The former yields the well known factor $V_{D-2}/(4\pi
s)^{D/2-1}$, whereas the latter has been computed in Ref.\
\cite{dow77} and is given by
\begin{equation}\label{hkcon}
\zeta_\beta(s)={\beta\over 2\pi}{A\over 4\pi
s}-
{1\over 4\pi s}\int_0^{\infty}d\xi\;
\xi \int_{-\infty}^\infty dw
e^{-\xi^2\cosh^2(w/2)/s} \cot{\pi\over\beta}(\pi+iw)
\end{equation}
The integrals can be done with the result that
\begin{equation}\label{hkcon2}
\zeta_\beta(s)={\beta\over 2\pi}{A\over 4\pi
s}-  {1\over 12}\biggl({2\pi\over \beta}-{\beta\over
2\pi}\biggr)
\end{equation}
with $A$ the area of the plane. At this point we must note that
when computing the free energy,
terms linear in $\beta$ in the heat kernel
correspond to zero-temperature contributions,
$F(\beta)=F^0+O(\beta^{-1})$ which, since they do not affect the
thermal behavior, should be subtracted (they are automatically
eliminated when computing the entropy using
Eq.\ (\ref{entf})). With these subtractions, the free energy is
proportional to the transverse space `area' and does not vanish
for $\beta=2\pi$. Taking all this into account we eventually find
the string free energy as
\begin{equation}\label{fcone}
\beta
{}F(\beta)=-{V_{D-2}\over 2\beta}{\pi\over 6} \int_E {d^2\tau\over
\tau_2^2} \lambda(\tau) \tau_2
\end{equation}
where $\lambda(\tau)$ is given by Eq.\ (\ref{lambd}), and $E$ is
the strip $ -1/2<\tau_1<1/2,\, \tau_2>0$.

The integrand is not modular invariant because of
the factor $\tau_2$ but, following Ref.\ \cite{obr}, we can
formally perform a coset extension and rewrite the free energy as
an integral over ${\cal F}$ of modular invariant quantities,
\begin{equation}\label{intxi0}
\beta F=-{V_{D-2}\over 2\pi\beta}
\int_{\cal F} {d^2\tau\over\tau_2^2}
\lambda(\tau){\sum_{m,n}}'{\tau_2\over |m\tau+n|^{2}}
\end{equation}
Remarkably, the $\beta$ dependence in this expression
is the same as in Eq.\ (\ref{f2}), and this
raises some issues:

\begin{itemize}

\item[-]
It must be noted
that $\beta=2\pi$ does not correspond to a high
temperature, but this may not be a real problem since
in any case the horizon involves fundamental aspects of
strings. Note also that, since the {\it local} temperature
$T_\xi=(\xi\beta)^{-1}$ grows without limit as we approach the
horizon, a {\it local} Atick-Witten behavior could be expected
at distances very close to the horizon. However, the results
above point to a different effect, namely, an Atick-Witten
dependence on the {\it global} temperature. As discussed below,
local interpretations are problematic.

\item[-]
Although the orbifold calculation and the Hamiltonian method
presented above are both based on conical backgrounds, the
origin of the Atick-Witten dependence seems to be
different in both cases. Actually, the $\beta$ dependence in
Eq.\ (\ref{intxi0}) could be objected on the grounds that it
directly follows from the conical heat kernel, Eq.\
(\ref{hkcon2}), after subtraction of terms linear in $\beta$.
Apparently, there are no stringy features here, in contrast
to the stringy peculiarities of the orbifold technique. This
leads us to a digression on the distinction between particles and
strings as regards to the physics near horizons. The
plausibility of the argument will rely more on physical
arguments rather than on technical ones.

We
start by recalling that the temperature dependence expected for a
particle field theory is $\beta F\propto \beta^{1-D}$.
Apparently, the bidimensional character of the $\beta$
dependence comes from the factorization of Rindler space into
$({\rm cone})\times({\rm transverse\; space})$, so that the
latter factor cannot not influence the $\beta$ dependence.
However, as argued in Ref.\ \cite{yo}, the correct $\beta$
dependence for particles is recovered after imposing a cutoff
$\xi_0$ near the horizon, which alters the scaling properties of
the heat kernel in the global space. The
introduction of this cutoff is related to the expectation that
the field theory description has to be radically altered near
the horizon \cite{thoo}. However, when considering strings we do
not expect a more fundamental description to be necessary;
instead, we need a more complete (non perturbative)
understanding of the phase structure of string theory. This is
needed here because the divergences in Eq.\ (\ref{intxi0})
appear in the infrared region $\tau_2\rightarrow\infty$, where
they are related to the Hagedorn instability. It is widely
accepted that infrared problems, as opposed to ultraviolet ones,
do not point to new underlying degrees of freedom, but rather to
changes in the qualitative behavior of the system. If we had
tried to perform the analysis above (i.e., by not introducing a
horizon regularization) for particles, the divergences would be
present in the ultraviolet region $s\rightarrow 0$. In this
respect, we just want to point out that the remarks made in the
analysis of Atick and Witten can be applied here, the small
$\beta$ limit in Ref.\ \cite{ati} being replaced by the limit
$\xi_0\rightarrow 0$. In both cases we need non-perturbative
knowledge to solve the problem of singularities. Notice
that, according to the arguments above, the association
of the heat kernel on the whole cone to the
thermodynamics of Rindler space would be physically
sensible only for a putatively fundamental theory like
string theory. That this yields a dependence on
temperature that may be physically relevant is a
remarkable fact.

\item[-] We now come to the problem of the
interpretation in
the black hole context of this {\it global} Atick-Witten
dependence. Note that, since we have
integrated all distances up to the horizon, the free energy that
we have computed not only contains the contribution of high
local-temperature degrees of freedom. This suggests that Eq.\
(\ref{intxi0}) {\it does not measure the degrees of freedom in
the usual sense of local quantum field theory} and the
interpretation of the $\beta$ dependence is not clear at all.
This is somewhat reminiscent of the well-known difficulties in
understanding in a statistical mechanical way the
Bekenstein-Hawking entropy associated to black holes ---notice
that we could have equally rephrased Eqs.\ (\ref{f2}) and
(\ref{intxi0}) in terms of the entropy, which is perhaps a more
meaningful quantity. Although it may be that the nontrivial
common features that we have found point to some universal
property ---perhaps in a spirit similar to the extension of
thermodynamics to include black holes---, it is difficult to
imagine a {\it gedanken} experiment to test in which sense the
Atick-Witten dependence that we are discussing represents a low
number of degrees of freedom of the string.

\end{itemize}

At this point we would like to compare the result of Eq.\
(\ref{intxi0}) with another approach recently applied to string
theory in Rindler space \cite{bar1,bar2,dab2}. This makes use of
a heuristic `red-shift argument' and involves two
basic steps:

\noindent (i) Obtain local thermodynamical densities by dividing
by the volume extensive quantities in flat space like the free
energy or the entropy.

\noindent (ii) The basic assumption of the red-shift argument:
local densities in Rindler space are equal to the ones in
{\it thermal} flat space by an adequate red-shift of the local
temperature,
$\beta\rightarrow\sqrt{g_{00}}\beta=\xi\beta$.

With these premises, one can construct a free energy density in
Rindler space out of the one in flat space, and then obtain the
global free energy by integrating over Rindler coordinates.
A free energy of this kind was first obtained in Ref.\
\cite{bar2}, and it can be readily checked that {\it when the
$\xi$ coordinate is integrated from $0$ to $\infty$ one finds
the same result as Eq.\ (\ref{intxi0})}.

However, there are problems concerning the validity of points (i)
and (ii) above. The definition of local
densities for strings is not as easy as point (i) assumes,
since localization of strings presents serious obstacles
\cite{cas}: confining strings in a box either yields an
infinite result by breaking conformal invariance or
introduces highly non-local behavior in the form of
solitons wrapping around the box.

Additional problems concern point (ii).
We want to stress that (ii) is just a heuristic
assumption, and it is important to notice that this
assumption on the {\it local} behavior of densities does
not follow from the {\it global} thermal character of the
Hartle-Hawking state.  What we actually know is that
Green's functions in Rindler space at temperature $1/2\pi$
are the same as the corresponding Green's functions of the
Minkowski observer at $T=0$. By no means it is immediate
to conclude from here the local behavior of densities
that the red-shift argument assumes.

These problems concerning the interpretation in terms of local
quantities can be conveniently
illustrated by noting that the red-shift arguments lead to a
determination of the position of
the local Hagedorn instability different from the one obtained
from the conical approach above.

In order to find the position of the local singularity predicted
by the conical approach we will consider the representation of
the free energy in terms of the integral over the strip $E$,
Eq.\ (\ref{hkfreest}). It is well known \cite{pol,obr} that in
the $E$ representation of the free energy the Hagedorn
singularity appears in the region $\tau_2\rightarrow 0$ (i.e.,
$s\rightarrow 0$). A close examination of this limit in the
$\xi$-dependent integrand of the heat kernel in Eq.\
(\ref{hkcon}),  together with the asymptotics
$|\eta(\tau)|^{-48}\sim \exp(4\pi/\tau_2)$ would
place the singularity at
$\xi_{Hag}=2\pi\sqrt{\alpha'}=\beta_c/2$. This is the value
previously found in Ref.\ \cite{par} by using a different
approach.

In contrast, the heuristic red-shift arguments would place the
Hagedorn transition at the point where the local
inverse temperature $\beta\xi$ equals the critical value
$\beta_c$, i.e., at $\xi'_{Hag}=\beta_c/\beta$ \cite{bar2,dab2},
clearly different from the previous value.

Therefore,
although the red-shift arguments may eventually lead to correct
global results, the underlying assumptions on the local behavior
are quite dubious. An attempt to give a formal basis for the
red-shift results can be found in Ref.\ \cite{bar2} where,
using a WKB approximation, a derivation of the red-shifted free
energy is given. In any case, due to the
extended character of the strings it is probably meaningless to
refer to any local behavior, a feature strongly suggested by the
`stretching effect' of strings near the horizon \cite{sus}. We
stress that neither the orbifold nor the Hamiltonian approach
presented in this paper rely on local properties.

There are a number of points that have been left out throughout
this work: (i) Interactions have been neglected. Their
importance to the problem of Hawking radiation has been
recently stressed in Ref.\ \cite{sus2}. It remains to be seen
how this influences the calculation of thermodynamical
quantities. (ii) In Ref.\ \cite{ati} it has been argued that the
Hagedorn singularity in thermal flat space at genus one means
that we have not treated correctly the genus zero contribution.
However,
an effective field theory argument has been given in
Ref.\ \cite{low} that such genus zero condensate vanishes (to
first order) in the Rindler background when $\beta=2\pi$.

In this paper we have given some tentative arguments pointing to
a behavior of the free energy of
strings in the presence of black hole horizons similar to the
high-temperature phase of string theory in flat space. In
both cases, the resolution of infrared singularities seems to
require non-perturbative knowledge of string theory.
However, although
independent orbifold and Hamiltonian approaches point to a
similar behavior, they can not be considered as rigorous
derivations and a sounder basis (e.g., by using an (unknown)
proper off-shell formulation) would be required. Also, it seems
difficult to interpret these results in terms of a counting
of degrees of freedom, a situation somewhat
reminiscent of well-known problems regarding the
statistical interpretation of the Bekenstein-Hawking entropy.
Finally, we have pointed out the problems of
using arguments based on local properties to study
thermodynamics in the presence of horizons.

We have greatly benefitted from useful criticism, advice and
encouragement from J.\ L.\ Ma\~nes. We are also indebted to
J.\ L.\ F.\ Barb\'on for helpful discussions on strings in
black hole backgrounds. This work has been partially supported by
a FPI grant from MEC (Spain) and projects UPV 063.320-EB119-92
and CICYT AEN93-0435.



\begin{thebibliography}{99}
\frenchspacing

\bibitem{gros} D. J. Gross and P. F. Mende, Nucl. Phys. B{\bf
303} (1988) 407.

\bibitem{ati} J. J. Atick and E. Witten, Nucl. Phys.
B{\bf 310} (1988) 291.

\bibitem{thoo} G. 't Hooft, Nucl. Phys. B{\bf 256} (1985)
727.

\bibitem{pol} J. Polchinski, Commun. Math. Phys. {\bf 104}
(1986) 37.

E. \'Alvarez and M. A. R. Osorio, Phys.
Rev. D{\bf 36} (1987) 1184.

\bibitem{obr} K. H. O'Brien and C. -I. Tan, Phys.
Rev. D{\bf 36} (1987) 1175.

B. McLain and B. D. B. Roth, Commun. Math. Phys. {\bf 111}
(1987) 539.

\bibitem{dix} L. Dixon, J. Harvey, C. Vafa and E.
Witten, Nucl. Phys. B{\bf 261} (1985) 678.

\bibitem{dab} A. Dabholkar, {\it Strings on a cone and black
hole entropy}, Harvard preprint HUTP-94-A019; hep-th/9408098.

\bibitem{low} D. A. Lowe and A. Strominger, {\it Strings
near a Rindler or black hole horizon}, Santa Barbara preprint
UCSBTH-94-42; hep-th/9410215.

\bibitem{bar2} J. L. F. Barb\'on, Phys. Lett. B{\bf
339} (1994) 41.

\bibitem{dow77} J. S. Dowker, J. Phys. A{\bf 10} (1977) 115.

\bibitem{yo} R. Emparan, {\it Heat kernels and
thermodynamics in Rindler space}, preprint EHU-FT-94/5;
hep-th/9407064.

\bibitem{bar1} J. L. F. Barb\'on, Phys. Rev. D{\bf
50} (1994) 2712.

\bibitem{dab2} A. Dabholkar, {\it Quantum corrections
to black hole entropy in string theory}, Caltech
preprint CALT-68-1953; hep-th/9409158.

\bibitem{cas} A. Casher, E. G. Floratos and N. C.
Tsamis, Phys. Lett. B{\bf 199} (1987) 377.
\bibitem{par} R. Parentani and R. Potting, Phys. Rev.
Lett. {\bf 63} (1989) 945.

\bibitem{sus}  L. Susskind, Phys. Rev. D{\bf 49} (1994)
6606.

\bibitem{sus2} L. Susskind and J. Uglum, {\it Black holes,
interactions and strings}, Stanford preprint SU-ITP-94-35;
hep-th/9410074.

\end{thebibliography}
\end{document}